\documentclass[prl,twocolumn,showpacs,superscriptaddress,nofootinbib]{revtex4-2}
\usepackage{graphicx}
\usepackage{braket}
\usepackage{amsmath,amsfonts,amssymb,amstext,amscd,amsthm,bbold}
\usepackage{comment,float}

\usepackage[dvipsnames]{xcolor}
\definecolor{myblue}{HTML}{03045E}
\usepackage[normalem]{ulem}
\usepackage{mathtools}

\usepackage{lipsum}
\usepackage{dcolumn}
\usepackage{bm}
\usepackage{verbatim}

\usepackage{orcidlink}
\hypersetup{
	colorlinks=true,
	citecolor= myblue,
	linkcolor=myblue,
	filecolor=myblue,      
	urlcolor=myblue,
	pdftitle={Overleaf Example},
	pdfpagemode=FullScreen,
}

\usepackage{titlesec}
\titleformat{\section}
{\sffamily\bfseries} % Set the sans-serif font here (\sffamily) and make it bold (\bfseries)
{\thesection} % The section number
{1em} % Space between the number and the heading
{} % Additional formatting if needed

\titleformat{\subsection}
{\sffamily\bfseries} % Set the sans-serif font here (\sffamily) and make it bold (\bfseries)
{\thesubsection} % The section number
{1em} % Space between the number and the heading
{} % Additional formatting if needed

\usepackage[T1]{fontenc}
\usepackage[utf8]{inputenc}

\titlespacing\section{0pt}{12pt plus 4pt minus 2pt}{2pt plus 2pt minus 2pt}
\titlespacing\subsection{0pt}{12pt plus 4pt minus 2pt}{2pt plus 2pt minus 2pt}
\titlespacing\subsubsection{0pt}{12pt plus 4pt minus 2pt}{2pt plus 2pt minus 2pt}
\usepackage{lipsum} % Include the lipsum package

\makeatletter
\let\@afterindenttrue\@afterindentfalse
\makeatother

\begin{document}

% \preprint{APS/123-QED}

\title{\textsf{Speed limits to the growth of Krylov complexity in open quantum systems}}% Force line breaks with \\
\thanks{All authors contributed equally to this work.}%
\author{Aranya Bhattacharya$^{\orcidlink{0000-0002-1882-4177}}$}
	\email{aranya.bhattacharya@uj.edu.pl}
	\affiliation{Institute of Physics, Jagiellonian University, Lojasiewicza 11, 30-348 Kraków, Poland. }

\author{Pingal Pratyush Nath$^{\orcidlink{0000-0001-5311-7729}}$}
\email{pingalnath@iisc.ac.in}
	\affiliation{Centre for High Energy Physics, Indian Institute of Science, C.V. Raman Avenue, Bangalore 560012, India. }
\author{Himanshu Sahu$^{\orcidlink{0000-0002-9522-6592}}$}
	\email{himanshusah1@iisc.ac.in}
		\affiliation{Centre for High Energy Physics, Indian Institute of Science, C.V. Raman Avenue, Bangalore 560012, India. }
	\affiliation{Department of Instrumentation \& Applied Physics, Indian Institute of Sciences, C.V. Raman Avenue, Bangalore 560012, Karnataka, India. }
	\affiliation{Department of Physics, Indian Institute of Science, Bangalore - 560012, Karnataka, India.}

\begin{abstract}
Recently, the propagation of information through quantum many-body systems, developed to study quantum chaos, have found many application from black holes to disordered spin systems. Among other quantitative tools, Krylov complexity has been explored as a diagnostic tool for information scrambling in quantum many-body systems. We introduce a universal limit to the growth of the Krylov complexity in dissipative open quantum systems by utilizing the uncertainty relation for non-hermitian operators. We also present the analytical results of Krylov complexity for characteristic behavior of Lanczos coefficients in dissipative systems. The validity of these results are demonstrated by explicit study of transverse-field Ising model under dissipative effects.

\end{abstract}
\maketitle

\noindent \textit{Introduction:} In quantum systems, interactions propagate the initially localized information across the exponentially large degrees of freedom \cite{PRXQuantum.5.010201,doi:10.1126/science.abg5029,PhysRevLett.124.200504,lewis-swanDynamicsQuantumInformation2019}. This phenomenon, known as \textit{quantum scrambling}, is crucial for addressing diverse unresolved questions in physics, such as the fast-scrambling conjecture for black holes\,\cite{yasuhirosekinoFastScramblers2008a,lashkariFastScramblingConjecture2013a}, peculiarities in strange metal behavior\,\cite{PhysRevD.96.106008,PhysRevB.97.155117}, and phenomena related to many-body localization\,\cite{PhysRevB.105.104202,otoc_localized}. Central to understanding quantum scrambling is the concept of out-of-time-order correlators (OTOC)\,\cite{PRXQuantum.5.010201,swingleUnscramblingPhysicsOutoftimeorder2018a} that are used to identify an analog of the Lyapunov exponent for systems in the semi-classical limit or having a large number of local degree of freedom\,\cite{shenkerBlackHolesButterfly2014a,PhysRevLett.118.086801}, thereby providing a connection with classical chaos. This ``quantum Lyapunov exponent'' exhibits a universal upper bound, attained by black holes\,\cite{maldacenaBoundChaos2016b,PhysRevLett.70.3339} and intertwined with the eigenstate thermalization hypothesis\,\cite{PhysRevLett.123.230606,PhysRevE.50.888}.

In this Letter, we consider another quantitative measure of quantum scrambling --- Krylov complexity\,\cite{PhysRevX.9.041017,barbonEvolutionOperatorComplexity2019a,rabinovici_operator_2021}. Krylov complexity (K-complexity) is a measure of the delocalization of a local intial operator evolving under Heisenberg evolution with respect to the Hamiltonian\,\cite{PhysRevX.9.041017,barbonEvolutionOperatorComplexity2019a,rabinovici_operator_2021,PhysRevB.102.085137,jianComplexityGrowthOperators2021,Bhattacharya_nonlocal}. It is conjectured to grow at most exponentially in non-integrable systems\,\,\cite{PhysRevX.9.041017} and can be used to extract the Lyapunov exponent, thereby, establishing a connection with OTOC\,\cite{maldacenaBoundChaos2016,PhysRevResearch.2.043234}. In isolated systems, a fundamental and ultimate limit to the growth of the K-complexity is introduced by formulating a Robertson uncertainty relation, involving the K-complexity operator and the Liouvillian, as generator of time evolution\,\cite{hornedalUltimateSpeedLimits2022b}. Such a bound is saturated by quantum systems in which the Liouvillian satisfies $\text{SU}(2)$, $\text{SL}(2,\mathbb{R})$ and the Heisenberg and Weyl algebra (HW) \cite{Caputa:2021sib}. %\textcolor{blue}{(which algebra do you mean? Or do you mean the recursion relation?)}.
These algebras arises naturally in certain quantum chaotic systems, such as the SYK model, but other chaotic systems do not maximize the growth of K-complexity.

Recently, the study of K-complexity has been extended to open quantum systems in which the operator growth is  governed by the Lindblad master equation\,\cite{Bhattacharya:2022gbz,Bhattacharya:2023zqt,PhysRevResearch.5.033085,Bhattacharjee:2022lzy,mohan2023krylov, Bhattacharjee:2023uwx}. In such systems, the information is generally lost to the environment which is reflected by the late time decay of K-complexity. In this Letter, we propose a fundamental speed limit to the growth of K-complexity in open quantum systems interacting with a Markovian bath. Since the operator evolution in open quantum systems is non-unitary, we employ the uncertainty relation for non-hermitian operators, thereby, obtaining a bound which depends on the \textit{probability decay}. The probability describes the loss of information to the environment, and the bound reduces to the closed system case in the absence of this term. In addition, we also give the analytical results of the growth of K-complexity in presence of purely imaginary diagonal Lanczos coefficients, which is a characteristics of open system as discussed in \cite{Bhattacharya:2023zqt}.

\smallskip 
%\vspace{5mm}
\noindent \textit{Brief survey of K-complexity in closed systems:} In an isolated system, the evolution of any operator $\mathcal{O}_0$ under a time-independent Hamiltonian $H$ is described by the Heisenberg equation of motion,
\begin{equation}\label{eq:operator_evolution}
\mathcal{O}(t) = e^{itH}\mathcal{O}_0 e^{-itH} = e^{i\mathcal{L}_ct} O_0 = \sum_{n=0}^\infty \frac{(it)^n}{n!}\mathcal{L}^n_c \mathcal{O}_0,
\end{equation}
where $\mathcal{L}_c$ is Hermitian Liouvillian superoperator given by $\mathcal{L}_c = [H,\,\bullet\, ]$. The operator $\mathcal{O}(t)$ can be expressed as a span of the nested commutators with the initial operator due to the repeated action of the Liouvillian as shown in Eq \eqref{eq:operator_evolution}. One constructs an orthonormal basis $\{|O_n)\}_{n=0}^{K-1}$ from this nested span of commutators, by choosing a certain scalar product $(\cdot|\cdot)$ on operator space. This orthogonal basis is known as the \textit{Krylov basis} and is achieved with the Lanczos algorithm --- a three term recursive version of the Gram-Schmidt orthogonalization method.

The dimension of Krylov space $K$ obeys a bound $K \leq D^2-D+1$, where $D$ is the dimension of the state Hilbert space\,\cite{rabinovici_operator_2021}. In the orthonormal basis $\{|O_n)\}$, the Liouvillian takes the tridiagonal form  $\mathcal{L}_c|O_n) = b_{n+1}|O_{n+1}) + b_n |O_n)$, where $b_n$ are referred to as Lanczos coefficients. The values $b_n$ are generated during the iterative steps of the orthogonalization process, signifying the characteristics of the scrambling process and serving as an indicator of chaos.

Once, the orthonormal basis is established, we can write the expansion of the operator $\mathcal{O}(t)$ as
\begin{equation}
\mathcal{O}(t) = \sum_{n = 0}^{K-1} i^n \phi_n(t) |O_n)
\end{equation}
The amplitudes $\phi_n(t)$ evolve according to the recursion relation $\partial_t\phi_n(t) = b_{n-1}\phi_{n-1}(t) - b_n\phi_{n+1}(t)$ with the initial conditions $\phi_n(0)=\delta_{n,0}$\,. The Lanczos coefficients $b_n$ can be thought of as hopping amplitudes for the initial operator $\mathcal{O}_0$ localized at the initial site to explore the \textit{Krylov chain}. The increase in support of operator away from the origin in Krylov chain reflects the growth of complexity as higher Krylov basis vectors are generated. To quantify this, one defines the average position of the operator in Krylov chain --- called the Krylov complexity as 
\begin{equation}
C(t) = (\mathcal{O}(t)|\mathcal{K}|\mathcal{O}(t)) = \sum_{n=0}^{K-1} n|\phi_n(t)|^2 
\end{equation}
where $\mathcal{K} = \sum_{n=0}^{K-1}n|O_n)(O_n|$ is position operator in the Krylov chain. The growth of Krylov complexity obeys an upper bound given by,
\begin{equation}\label{eq:bound_isolated}
|\partial_t C(t)| \leq 2b_1 \Delta \mathcal{K},
\end{equation}
where the dispersion of the position operator $\mathcal{K}$ is defined as $(\Delta \mathcal{K})^2 =\langle \mathcal{K}^2\rangle -\langle \mathcal{K}\rangle ^2$. One can define a characteristic time scale $\tau_K = \Delta \mathcal{K}/|\partial_t C(t)$ to write an analogue of the Mandelstam-Tamm bound as $\tau_K b_1\geq 1/2$.

\smallskip 
%\vspace{5mm}
\noindent \textit{K-complexity in open-quantum systems:} In open systems where the system interacts with an environment with weak coupling (Markovian bath), the dynamics of any operator is described by the Lindblad master equation
\begin{equation}
\mathcal{L}_o[\bullet] = [H,\bullet] - i\sum_k \left[L^\dagger_k\bullet L_k - \frac{1}{2}\{L^\dagger_k L_k,\bullet\}\right]
\end{equation}
where the operators $\{L_k\}$ are the Lindblad or the jump operators -- describes the nature of the interaction between the system and the environment. Since the Krylov basis $\{\mathcal{L}_o^n\mathcal{O}_0\}^{K-1}_{n=0}$ constructed from such a evolution in non-Hermitian, the usual Lanczos algorithm fails to orthonormalize them. Therefore, one resorts to alternatives such as Arnoldi or Bilanczos algorithms that are applicable to non-hermitian cases. In particular, the Bilanczos algorithm generates a bi-orthonormal basis $\{|p_n),|q_n)\}^{K-1}_{n=0}$ using the span $\{\mathcal{L}_o^n\mathcal{O}_0\}^{K-1}_{n=0}$ and $\{(\mathcal{L}_o)^\dagger \mathcal{O}_0\}^{K-1}_{n=0}$. These basis vectors obey the orthonormality relation $(q_m|p_n) = \delta_{mn}$. In such a basis, the non-Hermitian Lindbladian $\mathcal{L}_o$ can be written in a tridiagonal form 
\begin{align}\label{recurrence relations}
	c_{j+1} |p_{j+1}) &=\mathcal{L}_{o} |p_j) -a_j |p_j )  -b_{j} |p_{j-1})\,  \\
	b^*_{j+1} |q_{j+1}) &= \mathcal{L}_{o}^\dagger |q_j) - a^*_j |q_j) -c^*_{j} |q_{j-1})\,.
\end{align}

\begin{figure}[t]
    \centering
    \includegraphics[width = 0.9\linewidth]{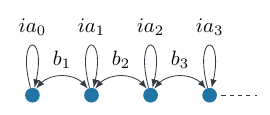}
    \caption{Schematic of Krylov chain for dissipative open systems in which hopping amplitudes between the sites are $b_n$ coefficients and on-site potential $ia_n$. }
    \label{fig:krylov_chain}
\end{figure}
The bra and ket versions of the time-evolved operator $\mathcal{O}(t)$ can, therefore, be expanded as 
\begin{equation}\label{operator expansion}
	\begin{split}
|\mathcal{O}(t)\rangle &= \sum_n i^n \phi_n(t) |p_n ), \\ 
\langle \mathcal{O}(t)| &= \sum_n (-i)^n \psi^*_n (t) ( q_n|.
	\end{split}
\end{equation}
The amplitudes $\phi_n(t)$ and $\psi_n(t)$ evolve according to the recursion relation 
\begin{equation}
	\label{differential equation for coefficients}
	\begin{split}
		\dot{\phi}_n(t) &= ia_n \phi_n - b_{n+1}\phi_{n+1} + c_n \phi_{n-1} \\
		\dot{\psi}^*_n(t) &= -ia^*_n \psi^*_n - c^*_{n+1}\psi^*_{n+1} + b^*_n \psi^*_{n-1}
	\end{split}
\end{equation}
with the initial conditions $\phi_n(0) = \psi_n(0 ) = \delta_{n,0}$\,. The numerical investigation in ref.\,\cite{Bhattacharya:2023zqt} shows that in open-quantum systems, the coefficients $a_n,b_n$ and $c_n$ obeys $b_n = c_n = |b_n|$ and $a_n = i|a_n|$, therefore, in what follows, we assume this to be valid. With this, the recursion relation for the amplitudes becomes $\dot{\phi}_n(t) = -|a_n| \phi_n - |b_{n+1}|\phi_{n+1} + |b_n| \phi_{n-1}$ and $\psi_n(t) = \phi_n(t)$\,. Therefore, in open systems, in addition to hopping amplitude $b_n$, there exist additional on site potentials $-|a_n|$ (See Fig.~\ref{fig:krylov_chain}). The purely imaginary nature of these on-site potentials result in decay of K-complexity showing the loss of information to environment. The K-complexity in analogy to isolated system case can be treated as
\begin{equation}
C(t) = \sum_{n=0}^{K-1} n\psi_n^*(t) \phi_n(t) = \sum_{n=0}^{K-1} n|\phi_n(t)|^2\,.
\end{equation}
 and also define the complexity operator in using the bi-orthogonal basis $|p_n), |q_n)$ as, 
 \begin{equation}
     \label{eq:complexityoperator}
     \mathcal{K} = \sum_{n =1}^{K-1} n|p_n)(q_n|\,.
 \end{equation}
% \vspace{5mm}
\textit{Thermodynamic limit:} In the continuum limit of $n$, we can write the recursion relation for $\phi_n(t)$ as
\begin{equation}
	\begin{split}
		\partial_t \phi(x,t) &= -a(x)\phi(x,t) -\partial_x b(x) \cdot \phi(x,t) \\
		& \qquad \qquad \qquad -2b(x) \cdot \partial_x \phi(x,t)\,.
	\end{split}
\end{equation}
We can make further simplification by making the substitution $b(x) \partial_x = \partial_y$ and $\chi(y,t) = \sqrt{b(x)}\phi(x,t)$ which leads to 
\begin{equation}\label{eq:partial_PDE}
    2\partial_y \chi_y(y,t) + \partial_t \chi(y,t) +\Tilde{a}(y) \chi(y,t) = 0
\end{equation}
where $\Tilde{a}(y) = a(x(y))$. The initial condition requires $|\phi(x,0)|^2 = \delta(x) $ which can also be translated to $|\chi(y,0)|^2 = b(x) \delta(x)$. The Eq.~\eqref{eq:partial_PDE} belongs to the generic family of first-order partial differential equations 
\begin{equation}\label{eq:PDE}
  f \partial_u \xi(u,v) + g \partial_v \xi(u,v) + q(u,v) \xi(u,v) = F(u,v)
\end{equation}
where $f,g$ are constants. The PDE~\eqref{eq:PDE} can be solved using suitable choice of characteristic curves\,\cite{ARFKEN2013401}, therefore, analytical result for the wave-function $\phi(x,t)$ can be found. The K-complexity $C(t)$ and total probability $P(t)$ defined in continuum as 
\begin{equation}
    C(t) = \int dx x |\phi(x,t)|^2\,;\quad  P(t) = \int dx |\phi(x,t)|^2\,.
\end{equation}
For few common choices of $a(x)$ and $b(x)$, the analytical results are\,\cite{supp} listed in Eq.~\eqref{eq:C_analytic}.

\begin{figure}
    \centering
    \includegraphics[width = 0.49\linewidth]{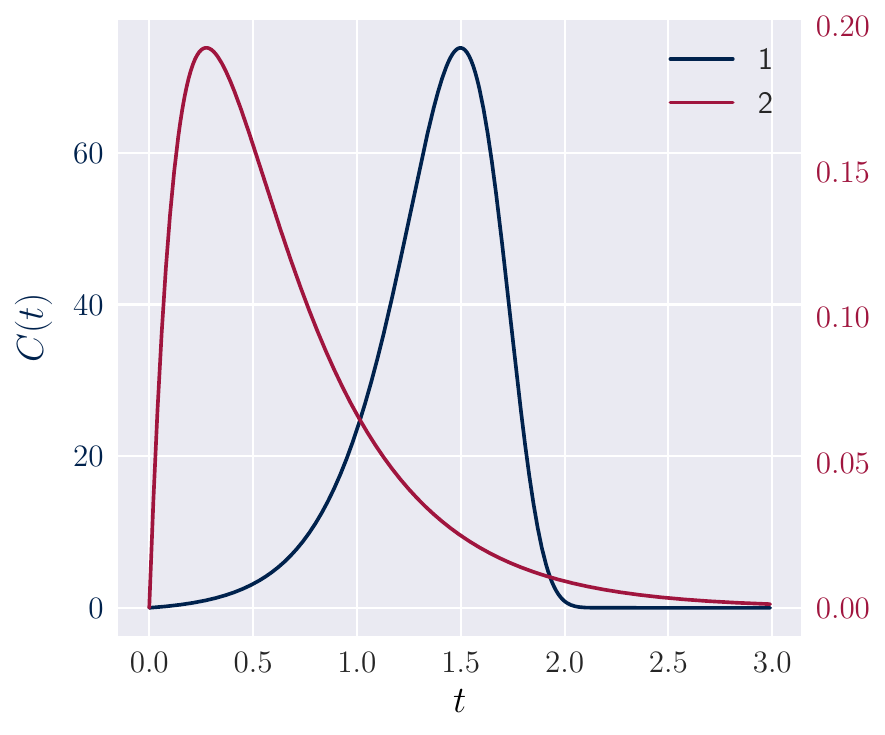}
    \includegraphics[width = 0.49\linewidth]{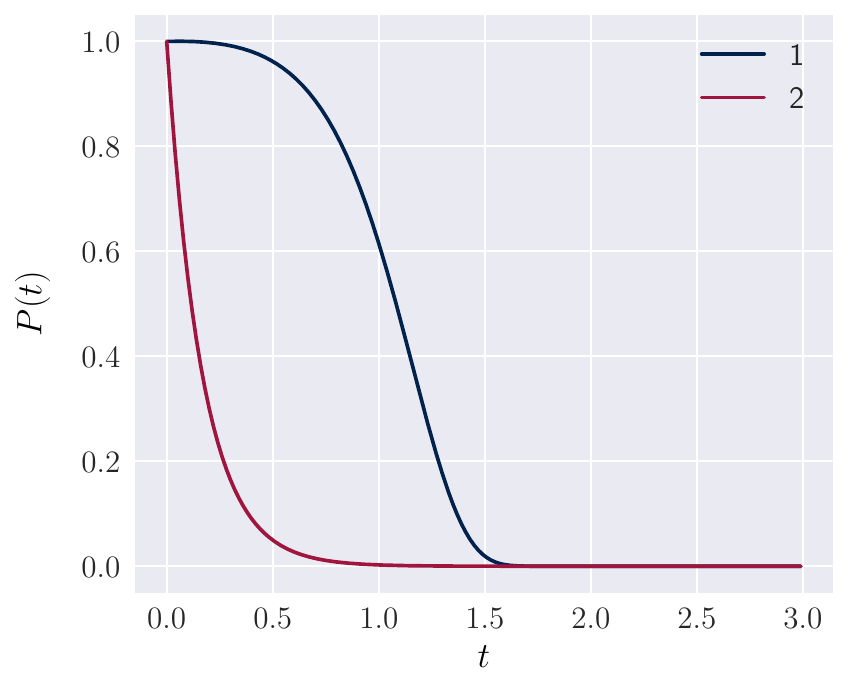}
    \caption{The analytic results of K-complexity $C(t)$ and the total probability $P(t)$ for two different choices (labeled as 1 and 2) of function $a(x)$ and $b(x)$ given in Eq.~\eqref{eq:C_analytic}. The parameter $(\alpha,\beta)$ for three choices are $\{(0.01,2),(3,2)\}$, respectively. In both cases, the complexity exponentially decays to zeros at late times.}
    \label{fig:analytic}
\end{figure}

\begin{figure*}
    \centering
    \includegraphics[width = 0.31\linewidth]{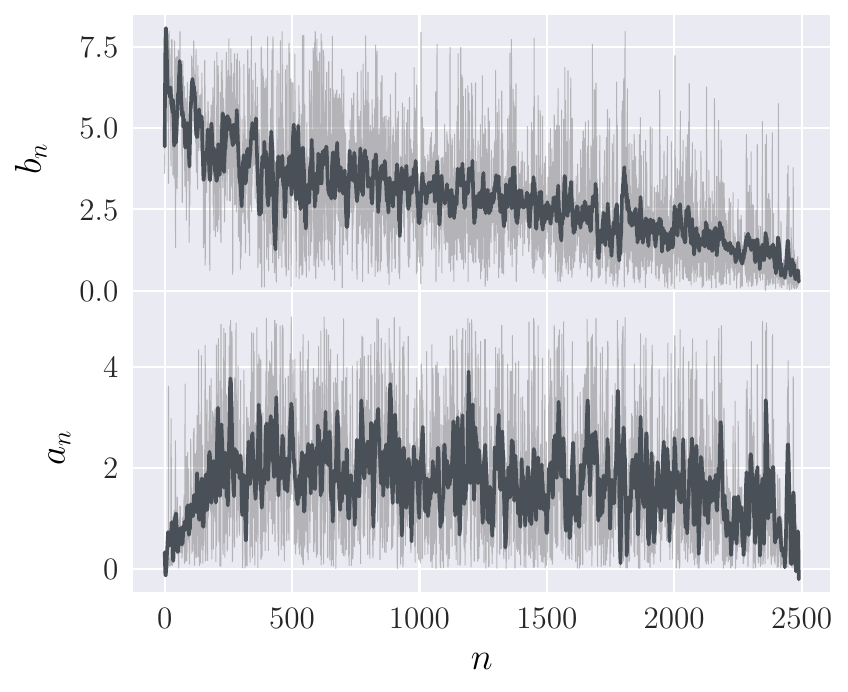}
    \includegraphics[width = 0.34\linewidth]{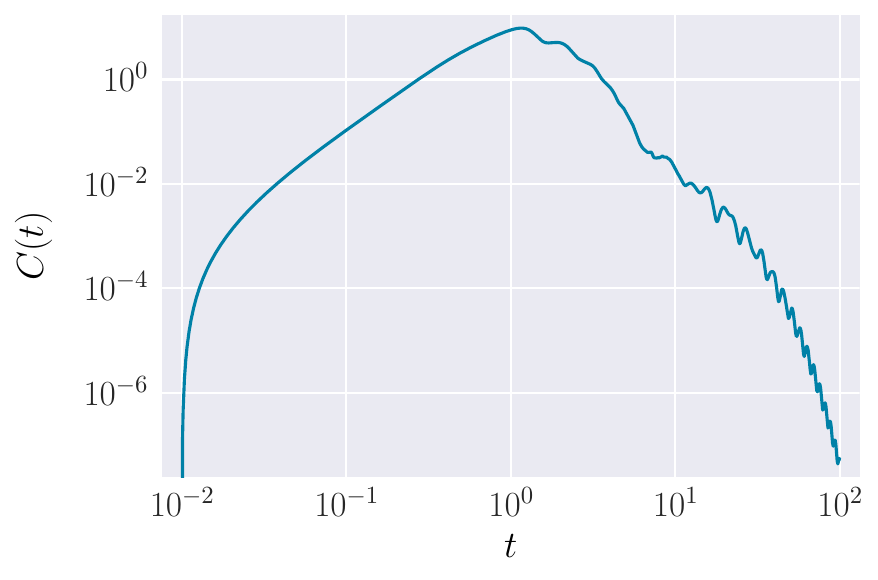}
    \includegraphics[width = 0.33\linewidth]{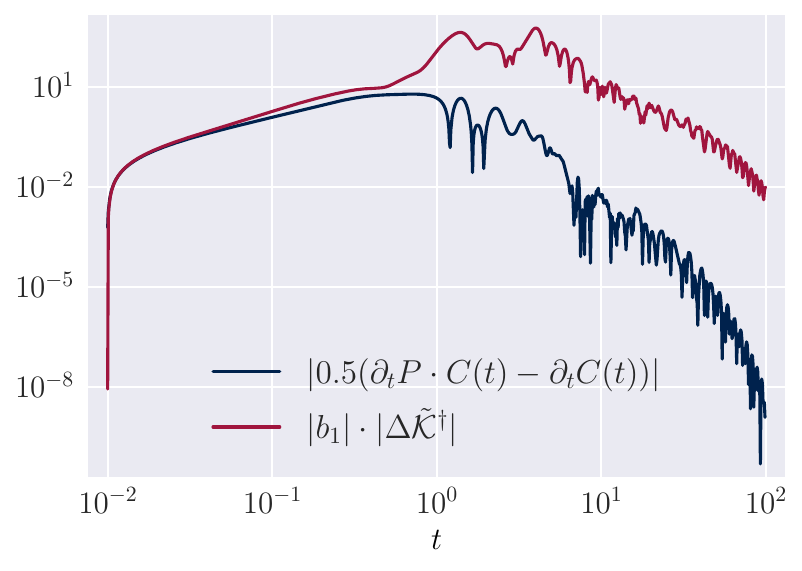}
    \caption{Growth of K-complexity in dissipative transverse-field Ising model with field coupling $(g,h)=(-1.05,0.5)$ and environment coupling $\alpha = \gamma  = 0.01$. \textbf{Left}: Lanczos coefficients (in light gray) $a_n, b_n$ after removing the outliers. The dark gray curve shows the averaged behavior obtained from filtering the original coefficients. \textbf{Center}: The K-complexity as a function of time $t$ in log-log plot. After the initial growth, the K-complexity decays to zero due to dissipation in system. \textbf{Right}: The illustration of dispersion bound in open-systems -- we show the left and right-hand side of inequality in Eq.~\eqref{eq:dispersion_bound}.}
    \label{fig:numerics}
\end{figure*}
The solution depicted in Figure \ref{fig:analytic} illustrates that, at late times, both the K-complexity and total probability exhibit exponential decay. The above choices %may appear arbitrary, nevertheless, 
are inspired by the numerical results of the growth of Lanczos coefficients in open systems and these capture various regimes of Lanczos coefficients\,\cite{Bhattacharya:2023zqt}. In thermodynamic limit, for open systems with boundary dephasing, $b_n$ will go through asymptotic linear growth\,\cite{Bhattacharya:2023zqt} while $a_n$ coefficients won't start growing at all. Hence, it will reduce to closed system dynamics and the corresponding speed limit holds. For bulk dephasing, $a_n$  grows from the beginning and the growth of complexity is similar to the first case we considered in Eq.~\eqref{eq:C_analytic}. 

In finite size system, the growth in $b_n$ is followed by a saturation, and the descent, while the growth in $a_n$ is followed by a saturation without showing any descent. The descent of $b_n$ features fluctuations which is large in integrable models compared to chaotic models. This results in suppression of saturation value in integrable model due to localization in the Krylov chain\,\cite{Rabinovici:2021qqt}.

\begin{widetext}
\begin{equation}\label{eq:C_analytic}
\begin{split}
C(t) &= 
     \begin{dcases}
       \frac{1}{\beta}\left( e^{2\beta t} -1\right)  \exp\left[\frac{2\alpha }{\beta } \left(  (1-e^{2\beta t})+5t \right)\right]  & \quad b(x) = \beta x + 1 \quad   \&  \quad a(x) = \alpha x; \\
       \frac{1}{\beta}\left(e^{2\beta t}-1\right) e^{-2\alpha t}  &  \quad b(x) = \beta x + 1 \quad   \&  \quad  a(x) = \alpha \,.
     \end{dcases} \\
P(t) &=
     \begin{dcases}
       \exp\left[\frac{2\alpha }{\beta } \left((1 - e^{-2\beta t})+5t \right)\right]  & \quad \qquad \qquad \quad  b(x) = \beta x + 1 \quad   \&  \quad a(x) = \alpha x; \\
       e^{-2\alpha t} &  \quad \qquad \qquad \quad  b(x) = \beta x + 1 \quad   \&  \quad  a(x) = \alpha \,.
     \end{dcases}
\end{split}
\end{equation}    
\end{widetext}
%\vspace{5mm}
\textit{Dispersion bound on K-Complexity in open-systems:} As we have seen, the operator evolution in open quantum systems is non-unitary, therefore, to consider the growth of K-complexity, we consider the uncertainty relation for non-hermitian operators. Apart from this, to consider the effect of the probability decay resulting from the non-hermiticity explicitly, we primarily frame the relation in terms of the un-normalized decaying complexity before recasting it in terms of the renormalized measures. We consider the uncertainty relation for non-hermitian operators $A$ and $B$ in a $d$-dimensional Hilbert space \cite{zhao_uncertainty_2022,PhysRevA.92.052120},
\begin{equation}\label{eq:non_her}
	\langle (\Delta A)^2\rangle \langle (\Delta B)^2\rangle \geq |\langle A^\dagger B\rangle -\langle A^\dagger \rangle \langle B\rangle|^2 
\end{equation}
where the variance of a non-Hermitian operator $O$ is defined as\,\cite{PhysRevA.92.052120}
\begin{equation}
	\langle (\Delta O)^2\rangle \equiv \langle O^\dagger O \rangle - \langle O^\dagger \rangle \langle O\rangle  .
\end{equation}
Using $A =  \tilde{\mathcal{K}}^\dagger \equiv \mathcal{K}/P(t)$ and $B =  \tilde{\mathcal{L}} =\mathcal{L}/P(t)$ --- normalized version of operators, and using the definitions of K-complexity, we can rewrite the uncertainty relation as\,\cite{supp}
\begin{equation}\label{eq:dispersion_bound}
	\left|\partial_tP(t) \cdot C(t) - \partial_t C(t) \right|^2 \leq 4|b_1|^2 (P(t))^2  \langle (\Delta \tilde{\mathcal{K}}^\dagger)^2 \rangle 
\end{equation}
where the expectation value taken with respect to operator $|\mathcal{O}(t))$. In terms of renormalized complexity defined as $\tilde{C}(t) = C(t)/P(t)$, we can recast the bound as 
\begin{equation}
\begin{split}
       & \left|(1-P(t))\cdot \partial_t P(t) \cdot \tilde{C}(t) + P(t)\cdot \partial_t \tilde{C}(t)\right|^2 \\
        & \qquad \qquad \qquad \qquad \leq 4|b_1|^2 \left(P(t)\right)^2 \langle (\Delta \Tilde{\mathcal{K}}^\dagger )^2\rangle \,.
\end{split}
\end{equation}
In isolated system, the total probability $P(t) = \sum_n |\phi_n(t)|^2$ is conserved so that $\partial_t P(t) = 0$ and $\tilde{\mathcal{K}}^\dagger = \mathcal{K}$. Therefore, the bound reduce to Eq.~\eqref{eq:bound_isolated} as expected.

In isolated systems, the dispersion bound is saturated iff the Lanczos coefficients grow according to\,\cite{hornedalUltimateSpeedLimits2022b}
\begin{equation}\label{eq:bn_saturate}
    b_n = \sqrt{\frac{1}{4} \alpha_0 n(n-1) + \frac{1}{2}\gamma_0 n}\,.
\end{equation}
For $\alpha_0 >1$ and large $n$, this reduces to linear growth $b_n  = \sqrt{\alpha_0}n$. In the thermodynamic limit, open systems under boundary dephasing alone, behaves similar to isolated systems since the seed operator is localized in the bulk and takes indefinite time to reach the boundary. Therefore, we expect the dispersion bound to be saturated for similar systems which satisfies Eq.~\eqref{eq:bn_saturate}.\footnote{In finite systems, the seed operator hit the boundary in at most scrambling time $t_s \sim \mathcal{O}(N)$, therefore, the results of isolated case are expected to hold for time smaller than $t_s$.} 

\vspace{5mm}
\noindent \textit{Numerical Results:} To illustrate the validity of the bound in Eq.~\eqref{eq:dispersion_bound}, we study the transverse-field Ising model Hamiltonian for $N$ spins, given by,
\begin{align}
	H = - \sum_{j=1}^{N-1}  \sigma^{z}_{j} \sigma^{z}_{j+1} - g \sum_{j=1}^{N} \sigma^{x}_j - h \sum_{j=1}^{N} \sigma^{z}_{j}\,, \label{tfim}
\end{align}
where $g$ and $h$ are the coupling parameters. The interaction with the environment are encoded in the set of Lindblad operators $L_k$ : 1) $\sqrt{\alpha} \sigma_k^{\pm}$ with $k \in \text{boundary}$ 2) $\sqrt{\gamma} \sigma_i^z$ with $k \in \text{bulk}$, where $\alpha,\gamma >0$ is the coupling strength between the system and the environment and $\sigma^{\pm}_k = (\sigma^x_k \pm i\sigma^y_k)/2$. 
For our numerical analysis, we will take field coupling as $g = -1.05$, $h  = 0.5$ and environmental coupling $\alpha = \gamma = 0.01$. We choose an initial observable to be uniformly distributed operator $(1/d,1/d,\ldots ,1/d)^T$. 
We utilize the vectorized form of the Lindbladian, expressed in terms of the Hamiltonian $H$ and the Lindblad operators $L_k$,
\begin{equation}
\begin{split}
        \mathcal{L}_o &= (I \otimes H - H^T \otimes I) + \frac{i}{2} \sum_k \left.[I \otimes L_k^{\dagger} L_k \right.\\ 
        & \qquad \qquad  \qquad \left.+ L_k^T L_k^{*} \otimes I - 2 \, L_k^{T} \otimes L_k^{\dagger} \right]\,
\end{split}
\end{equation}
where $k$ iterates over the Lindblad operators. We implement the bi-Lanczos algorithm, incorporating full orthogonalization twice within the process to ensure the establishment of an orthogonal basis.

The left panel of Fig.~\ref{fig:numerics} shows the lanczos coefficients $b_n$ and $a_n$ for system size $N = 6$. The center panel of Fig.~\ref{fig:numerics} shows the K-complexity which exhibits exponential growth follows by decay due to environmental coupling. The rightmost panel illustrates dispersion bound on K-complexity in open-systems (Eq.\eqref{eq:dispersion_bound}) in log-log plots. 
%\vspace{5mm}
\smallskip 
\noindent \textit{Conclusion:} Our results establish the ultimate speed limit to operator growth in open quantum systems. We showed that the dispersion bound and the wave function decay governs the complexity growth rate in most general versions of open system dynamics. This bound holds for both finite sized open systems and in the thermodynamic limit with both boundary and bulk dephasing. In \cite{PhysRevD.106.046007, Caputa:2024vrn}, the authors introduce an analogous notion of complexity for quantum many-body states, defined as a spread in the Krylov basis formed by the Hamiltonian of the system -- dubbed as the \textit{spread complexity}. The K-complexity dispersion bound for both isolated, open, and measurement-induced systems \cite{Bhattacharya:2023yec} can be extended to spread complexity. In this case, it is important to note the presence of Lanczos coefficients $a_n$ in both isolated and open cases. A extension of this work could explore the form of bound for the spread complexity.

Another interesting direction could be to consider the quantum-speed limit bound namely Mandelstam and Tamm (MT) bound and Margolus and Levitin (ML) bound\,\cite{Mandelstam1991,MARGOLUS1998188,Deffner_2017} for the operator evolving in the krylov chain. A key insight is to consider the operator-state mapping, usually known as ``Choi-Jamiolkowski isomorphism'' or channel-state duality. Under such operator-state mapping, the Lindbladian dynamics in operator space reduces to the Hamiltonian dynamics in state space with extended dimension. Therefore, analogous bound to MT (and ML) bounds can be derived. 

In such a dual space, the open system case corresponds to effective Hamiltonian of the form $\tilde{H}_\text{eff} = \tilde{H} - i\tilde{\Gamma}$ in Krylov basis where $\Tilde{H}$ is a tridiagonal matrix and $\Gamma$ is a diagonal matrix with matrix elements as $b_n$ and $a_n$, respectively.  Therefore, the speed limit bound provided in ref.~\cite{PhysRevLett.110.050403} should hold.

\smallskip 
%\vspace{5mm}

\noindent \textit{Acknowledgement:} We wish to thank Aninda Sinha and Aldolfo Del Campo for various useful discussions and comments about this and related works. The work of AB is supported by the Polish National Science Centre (NCN) grant 2021/42/E/ST2/00234.

\bibliographystyle{apsrev4-1}
\bibliography{ms}

\clearpage
\newpage 
\onecolumngrid

\renewcommand{\theequation}{S\arabic{equation}}
\renewcommand{\thefigure}{S\arabic{figure}}
\renewcommand{\bibnumfmt}[1]{[S#1]}
\renewcommand{\citenumfont}[1]{S#1}
\renewcommand\figurename{Supplementary Fig.}

\begin{center}
	\textbf{\large \textsf{Supplemental Materials: Speed limits to the growth of Krylov complexity in open quantum systems} }
\end{center}

\noindent \textbf{\textsf{Bilanczos Algorithm}}\label{app:bilanczosalgorithm} We outline the detailed algorithm to construct the orthonormal basis for operator complexity in open quantum systems and spread complexity in Non-Hermitian systems. The algorithm takes the initial operators/states $|p_0\rangle, |q_0\rangle$ and the Lindbladian $\mathcal{L}_{o}$ as inputs and outputs the coefficients $a_n$, $b_n$ and $c_n$ along with the basis $|p_n\rangle$ and $\langle q_n|$. 
%For the open system, we use the vectorized form of the Lindbladian as $\mathcal{L_0}$ and the initial operator as $O_0 = (1/d,1/d, \dots)^{T}$ of length $d = 4^N$, where $N$ is the number of spins of the system. For the Non-Hermitian case, we use the Non-Hermitian Hamiltonian as  $\mathcal{L_0}$, and the initial vector is the same as the one used in Operator complexity but with dimension $d = 2^N$. 

\vspace{2 mm}

\noindent The steps of the algorithm are as follows:

\begin{enumerate}
	\item Let $|p_0\rangle, |q_0\rangle \in \mathbb{C}^n$ be arbitrary vectors with $\langle q_0|p_0\rangle = 1$., i.e., we choose $|p_0\rangle = |q_0\rangle \equiv |O_0)$ at the initial step.
	\item The initial iteration steps are given as follows:
	\begin{enumerate}
		\item Let $|r'_0\rangle =\mathcal{L}_{o} |p_0\rangle$ and $|s'_0\rangle =\mathcal{L}_{o}^\dagger  |q_0\rangle$.
		\item Compute the inner product $a_0 = \langle q_0|r'_0 \rangle$.
		\item Define $|r_0\rangle =|r_0'\rangle -a_0 |p_0\rangle$ and $|s_0\rangle = |s'_0\rangle -a^*_0 |q_0\rangle$.
	\end{enumerate}
	
	\item for $j=1,2, \ldots$, perform the following steps: 
	\begin{enumerate}
		\item Compute the inner product $\omega_{j}=\langle r_{j-1}|s_{j-1} \rangle$.
		\item Compute the norm $c_{j}= \sqrt{|\omega_{j}|}$ and $b_{j} = \omega^*_{j}/c_{j}$.
		\item If $c_{j}\neq0$, let 
		\begin{align}
			|p_j\rangle = \frac{|r_{j-1}\rangle}{c_{j}}~~~~ \& ~~~~ |q_j\rangle =\frac{|s_{j-1}\rangle}{b^*_{j}}\,.
		\end{align}
		\item If required, perform the full orthogonalization
		\begin{align*}
			|p_j\rangle &= |p_j\rangle - \sum_{i=0}^{j-1} \langle q_i|p_j \rangle\, |p_i \rangle\,, \\
			|q_j\rangle &= |q_j\rangle - \sum_{i=0}^{j-1} \langle p_i|q_j \rangle \, |q_i \rangle\,.
		\end{align*}
		\item Let $|r'_j \rangle = \mathcal{L}_{o} |p_j \rangle $  and $|s'_j\rangle = \mathcal{L}_{o}^\dagger  |q_j\rangle$. 
		\item Compute $a_{j} = \langle q_j|r'_j \rangle$.
		\item Define the vectors:
		\begin{align*}
			|r_j\rangle &= |r'_j \rangle  -a_{j}|p_j\rangle -b_{j} |p_{j-1}\rangle\,, \\
			|s_j\rangle &= |s_j'\rangle -a^*_{j} |q_j\rangle -c^*_{j}|q_{j-1}\rangle\,. 
		\end{align*}
		and go back to step $3$.
	\end{enumerate}
	\item If $c_j=0$ for some $j = \mathcal{K}-1$, where $\mathcal{K}$ is the Krylov dimension, let $P= (p_0~\,p_1 ~\,\ldots~\,p_{\mathcal{K}-1})$, and $Q= (q_0~\,q_1 ~\,\ldots~\,q_{\mathcal{K}-1})$. The Lindbladian is then given by $[\mathcal{L}_o]=Q^{\dagger} \mathcal{L}_o P$.
\end{enumerate}

\noindent \textbf{\textsf{Exact solutions of Krylov complexity}}\label{app: exact solutions of K-complexity} The differential recurrence relation with the assumption $b_n = c_n$ and $a_n = i|a_n|$, simplifies to
\begin{equation}
	\dot{\phi}_n(t) = -|a_n| \phi_n - b_{n+1} \phi_{n+1} + b_n \phi_{n-1}
\end{equation}
In the continuous limit of $n$, we can rewrite this as \footnote{Note that our exact differential equation is a bit different from the approximate version considered in section 6.1 of \cite{Bhattacharjee:2022lzy} which results in late-time value of the wavefunctions having power-law
	growth in x in addition to the
	exponential suppression.} 

\begin{equation}
	\begin{split}
		\partial_t \phi(x,t) &= -a(x)\phi(x,t) -\partial_x b(x) \cdot \phi(x,t)  -2b(x) \cdot \partial_x \phi(x,t)
	\end{split}
\end{equation}
Using the substitution $b(x)\partial_x = \partial_{y}$ and $\chi(y,t) = \sqrt{b(x)} \phi(x,t) $, we have 

\begin{equation}\label{eq:diff}
	\partial_y \chi(y,t) = \frac{1}{2}\left[-\tilde{a}(y) \chi(y,t)-\partial_t \chi(y,t)  \right]
\end{equation}
where $\tilde{a}(y)  = a(x(y))$ along with initial condition $\chi(y,0) = \chi_i(y)$. The Krylov complexity in continuous limit can be rewritten as 
\begin{equation}
	C(t) = \sum_{n=0}^{D-1} n|\phi_n(t)|^2\rightarrow \int dx x |\phi(x,t)|^2\,.
\end{equation}

\hrule 
\vspace{5mm}

\noindent \textit{On solution method of PDE}-- The PDE in Eq.~\eqref{eq:diff} belongs to family of first-order PDE of a form 
\begin{equation}
	\mathcal{F} \varphi = a \partial_x \varphi + b \partial_y\varphi + q(x,y) \varphi = F(x,y)
\end{equation}
We can identify its characteristic curves, amounting to transformating to new set of variables $s = ax+by$, $t = bx-ay$, in terms of which our PDE becomes 
\begin{equation}
	(a^2 + b^2) \partial_s \varphi + \hat{q}(s,t) \varphi = \hat{F}(s,t)\,.    
\end{equation}
where 
$$\hat{q}(s,t) = q\left(\frac{ax+bt}{a^2 +b^2},\frac{bs-at}{a^2+b^2}\right)$$
and similar change of variables to write $\hat{F}$ from $F$. The above equation is an ODE in $s$ which can be solved using various method for solving ODEs.\\

\hrule 
\vspace{5mm}

\noindent To find the exact solution of Eq.~\eqref{eq:diff}, we will assume the explicit form of $a(x)$ and $b(x)$ in the thermodynamic limit. Note that the off-diagonal coefficients in the continuous limit ($b(x)$) can grow asymptotically forever. However, the initial growth of the diagonal coefficients in the continuous limit ($a(x)$) depends only on the coupling of the boundary dephasing and the growth stops once the effect of the interaction reaches a certain maximum limit beyond which it can not effect the system part anymore. Therefore, we consider the two following cases. i) linear growth of both $b(x)$ and $a(x)$. ii) Linear growth of $b(x)$ and saturated constant value of $a(x)$.

\noindent \textit{Case I} Consider $b(x) = \beta x + c$ and $a(x) = \alpha x$ so that 
\begin{equation}
	\begin{split}
		dx = b(x) dy =( \beta x +c) dy \rightarrow x = \frac{1}{\beta}(c_0e^{\beta y} -c )
	\end{split}
\end{equation}
where $c_0$ is a constant fixed by imposing the normalization at $t=0$. 
\begin{align*}
	|\phi(x,0)|^2 &= \delta(x) \\
	|\chi(y,0)|^2 &= b(x) \delta(x) = (\beta x + c) \delta(x) = c_0 e^{\beta y}\delta\left(\frac{1}{\beta}(c_0e^{\beta y}-c)\right) \\
	&= \frac{c}{c_0} e^{\beta y} \delta(y) \rightarrow \int dy |\chi(y,0)|^2 = \frac{c}{c_0} =1 \rightarrow c_0 = c
\end{align*}
so that 
\begin{equation}
	x = \frac{c}{\beta}( e^{\beta y} -1) \rightarrow y = \frac{1}{\beta}\ln \left(\frac{\beta x}{c}  + 1\right)
\end{equation}
and
\begin{equation}
	\tilde{a}(y) = a(x(y)) = \frac{\alpha c}{\beta} \left[e^{\beta y} - 1\right]    
\end{equation}
The equation \eqref{eq:diff} becomes 
\begin{equation}
	2\partial_y \chi(y,t) + \partial_t \chi(y,t) +\frac{\alpha c}{\beta} \left[e^{\beta y} - 1\right]      \chi(y,t) = 0
\end{equation}
Identifying the characteristic curves, the solution is found to be
\begin{equation}
	\chi(y,t) = \zeta (y-2t) \exp\left[-\frac{\alpha c}{\beta}\left(e^{\beta y}-(2y+t)\right) \right]
\end{equation}
Imposing the initial condition, we get
\begin{align*}
	e^{\beta y} \delta(y) &= |\zeta(y)|^2 \exp\left[-\frac{2\alpha c}{\beta}\left(e^{\beta y}-2y\right) \right] \\
	\rightarrow |\zeta (y)|^2 &= e^{\beta y}\delta(y) \exp\left[\frac{2\alpha c}{\beta}\left(e^{\beta y}-2y\right) \right]
\end{align*}
Therefore,
\begin{equation}
	|\chi(y,t)|^2 = e^{\beta (y-2t)}\delta(y-2t) \exp\left[\frac{2\alpha c}{\beta } \left(e^{\beta y} (e^{-2\beta t}-1)+5t \right)\right]
\end{equation}
The complexity given by 
\begin{align*}
	C(t) &= \int dy x(y) |\chi(y,t)|^2 = \frac{c}{\beta}\left( e^{2\beta t} -1\right)  \exp\left[\frac{2\alpha c}{\beta } \left(  (1-e^{2\beta t})+5t \right)\right]
\end{align*}
putting $c=1$ so that 
\begin{equation}
	C(t) = \frac{1}{\beta}\left( e^{2\beta t} -1\right)  \exp\left[\frac{2\alpha }{\beta } \left(  (1-e^{2\beta t})+5t \right)\right] 
\end{equation}
The normalization given by  
\begin{equation}
	\begin{split}
		P(t) &= \int dy |\chi(y,t)|^2 =  \exp\left[\frac{2\alpha }{\beta } \left((1 - e^{-2\beta t})+5t \right)\right]\,.
	\end{split}
\end{equation}
\noindent \textit{Case II} Consider  $ b(x) = \beta x + c$ and $a(x) = \alpha $ so that 
\begin{equation}
	\begin{split}
		dx &= b(x) dy = (\beta x +c)dy \rightarrow 
		y = \frac{1}{\beta}\ln \left[ -\frac{\beta x}{c}+1\right]
	\end{split}
\end{equation}
and $\tilde{a}(y) = a(x(y)) = \alpha $. The equation becomes 
\begin{equation}
	2\partial_y \chi(y,t) + \partial_t \chi(y,t) +\alpha  \chi(y,t) = 0
\end{equation}
Identifying the characteristic curves, the solution is found to be 
\begin{equation}
	\chi(y,t) = \zeta(y-2t) \exp\left[-\frac{\alpha}{5}(2y+t)\right]
\end{equation}
The initial condition requires 
\begin{align*}
	e^{\beta y} \delta(y) &= |\zeta(y)|^2 \exp\left[-\frac{4\alpha}{5}y\right]  \rightarrow |\zeta (y)|^2 = e^{\beta y}\delta(y)  \exp\left[\frac{4\alpha}{5}y\right]
\end{align*}
which in turn gives 
\begin{equation}
	|\chi(y,t)|^2 = e^{\beta (y-2t)}e^{-2\alpha t}\delta(y-2t)
\end{equation}
The complexity is given by 
\begin{align*}
	C(t) &= \int dy \ x(y) |\chi(y,t)|^2 = \frac{c}{\beta}\left(e^{2\beta t} -1\right) e^{-2\alpha t}
\end{align*}
Therefore, 
\begin{equation}
	C(t) =\frac{c}{\beta}\left(e^{2\beta t}-1\right) e^{-2\alpha t}
\end{equation}
The normalization is given by 
\begin{equation}
	P(t) = 	\int dy |\chi(y,t)|^2 = e^{-2\alpha t}.
\end{equation}
In summary, 
\begin{equation}\label{eq:C_analytic}
	\begin{split}
		C(t) &= 
		\begin{dcases}
			\frac{1}{\beta}\left( e^{2\beta t} -1\right)  \exp\left[\frac{2\alpha }{\beta } \left(  (1-e^{2\beta t})+5t \right)\right]  & \quad b(x) = \beta x + c \quad   \&  \quad a(x) = \alpha x;\\
			\frac{c}{\beta}\left(e^{2\beta t}-1\right) e^{-2\alpha t}  &  \quad b(x) = \beta x + c \quad   \&  \quad  a(x) = \alpha \,.
		\end{dcases} \\
		P(t) &=
		\begin{dcases}
			\exp\left[\frac{2\alpha }{\beta } \left((1 - e^{-2\beta t})+5t \right)\right]  &  \qquad \qquad \quad  b(x) = \beta x + c \quad   \&  \quad a(x) = \alpha x;\\
			e^{-2\alpha t} & \qquad \qquad \quad  b(x) = \beta x + c \quad   \&  \quad  a(x) = \alpha \,.
		\end{dcases}
	\end{split}
\end{equation} 
\noindent \textbf{\textsf{Derivation of dispersion bound on Krylov complexity}}\label{app:derivation_of_dispersion_bound} For the choice of $A =  \mathcal{K}^\dagger $ and $B =  \mathcal{L}$, the uncertainty relation for Non Hermitian operators introduced in the main text, reads 
\begin{equation}\label{eqsup:uncertainityrelation}
	\langle (\Delta \mathcal{K}^\dagger)^2\rangle \langle (\Delta \mathcal{L})^2\rangle \geq |\langle \mathcal{K}\mathcal{L}\rangle -\langle \mathcal{K} \rangle \langle \mathcal{L}\rangle|^2 
\end{equation}
Although, since we are dealing with open-quantum systems, it is more nature to work with the renormalized version of these operators \textit{i.e.} $\Tilde{\mathcal{K}} = \mathcal{K}/P(t)$ and $\Tilde{\mathcal{L}} = \mathcal{L}/P(t)$. We will employ the expansion of the operator $\ket{O(t)}$ (and $\bra{O(t)}$) in terms of basis elements $|p_n\rangle$ (and $\langle q_n|$), along with the action of the Lindbladian $\mathcal{L}$ om these basis elements as outlined in the bilanczos algorithm. This approach allows us to obtain the different terms in the uncertainty relation mentioned above.

\vspace{2 mm}
\noindent $\langle \mathcal{L}\rangle$ : 
\begin{align*}
	\mathcal{L}|\mathcal{O}(t)) &= \sum_n i^n \phi_n(t)\mathcal{L}|p_n\rangle \\
	&= \sum_n i^n \phi_n(t) \left[a_n|p_n\rangle + b_n |p_{n-1}\rangle + c_{n+1}|p_{n+1}\rangle \right] \\
	\langle \mathcal{L}\rangle  &= \sum_{n,m} i^n (-i)^m \phi_n \psi^*_m \left[a_n \delta_{n,m} + b_n \delta_{n-1,m} \right.  \left.  + c_{n+1}\delta_{m,n+1} \right] \\
	&= \sum_n \phi_n (t) \left[a_n \psi^*_n + ib_n \psi^*_{n-1} -ic_{n+1}\psi^*_{n+1} \right] \\
	&= \sum_n \psi^*_n(t) \left[a_n \phi_n(t) + ib_{n+1}\phi_{n+1}-ic_n \phi_{n-1} \right] = -i \sum_n \psi^*_n(t) \dot{\phi}_n(t)
\end{align*}
In the final step, we applied the recursion relation for $\phi_n$ in the bilanczos algorithm to simplify the expression. Consequently, the expectation value of $\mathcal{L}$ can be expressed as follows,	
\begin{equation}
	\langle \mathcal{L}\rangle = -i \sum_n \psi^*_n(t) \dot{\phi}_n(t)
\end{equation}
We utilize the Complexity operator in terms of the basis elements $\bra{p_n}$ and $\ket{q_n}$ to compute the expectation value of $\mathcal{K}$, which simplifies to the expression of Krylov Complexity for the bilanczos algorithm. 

\noindent $\langle\mathcal{K}\rangle$ : 
\begin{align*}
	\mathcal{K} |\mathcal{O}(t)) &= \sum_n n |p_n\rangle \langle q_n | \left(\sum_m i^m \phi_m |p_m\rangle \right) \\
	&= \sum_n n i^n\phi_n |p_n\rangle \\
	\langle \mathcal{K} \rangle &= \sum_n n \phi_n \psi^*_n
\end{align*}
In a similar fashion, we calculate the expectation value of $\mathcal{K}\mathcal{L}$ by employing the definitions of the complexity operator and the action of the Lindbladian on the basis elements.

\noindent $\langle \mathcal{K} \mathcal{L}\rangle$ :              
\begin{align*}
	\mathcal{L}|O(t)) &= \sum_n i^n \phi_n(t)\mathcal{L}|p_n\rangle \\
	&= \sum_n i^n \phi_n(t) \left[a_n|p_n\rangle + b_n |p_{n-1}\rangle + c_{n+1}|p_{n+1}\rangle \right] \\
	\mathcal{K}  \mathcal{L}|O(t)\rangle &= \sum_{n,m}i^n\phi_n(t) m|p_m\rangle \left[c_{n+1}\delta_{m,n+1}\right.  \left. + a_n\delta_{m,n} + b_n \delta_{m,n-1}\right] \\
	&= \sum_n  i^n\phi_n(t) \left[n a_n |p_n\rangle + b_n (n-1)|p_{n-1}\rangle \right.  \left.  + (n+1) c_{n+1}|p_{n+1}\rangle  \right] 
\end{align*}    
We utilize the orthogonality relations of the basis elements, $\braket{p_n|q_n} = \delta_{mn}$, to simplify the above expression. Subsequently, we employ this simplification to obtain a concise expression for the expectation value of $\mathcal{K}\mathcal{L}$.
\begin{align*}
	\langle \mathcal{K}  \mathcal{L}\rangle &= \sum_{n,m}i^n (-i)^m \phi_n\psi^*_m \left[na_n \delta_{n,m} + b_n (n-1) \delta_{n-1,m} \right.  \left.  + c_{n+1}(n+1) \delta_{n+1,m} \right] \\
	&= \sum_n i^n \phi_n(t) \left[n a_n (-i)^n \psi_n^* \right. + (n-1)b_n (-i)^{n-1}\psi^*_{n-1}    + (n+1)c_{n+1}(-i)^{n+1}\psi^*_{n+1} ] \\
	&= \sum_n \phi_n \left[n a_n \psi^*_n + ib_n (n-1) \psi^*_{n-1} \right. \left.  -i (n+1) c_{n+1}\psi^*_{n+1}  \right] 
\end{align*}
Therefore
\begin{equation}
	\langle  \mathcal{K}  \mathcal{L}\rangle = \sum_n n\psi^*_n \left[a_n \phi_n + ib_{n+1} \phi_{n+1} - ic_n \phi_{n-1}\right]
\end{equation}
We subsequently apply the recursion relation for $\phi_n$ to further simplify this expression, and it can be written as,
\begin{equation}
	\langle  \mathcal{K}  \mathcal{L}\rangle = -i\sum_n n \psi^*_n(t) \dot{\phi}_n(t) .
\end{equation}
To further simplify, we consider $|b_n| = |c_n|$ and $a_n = i|a_n|$ which implies that $\psi_n(t) = \phi_n(t)$. Under these assumptions the simplified expectation values can be written as, 
\begin{equation}
	\begin{split}
		\langle  \mathcal{K}  \mathcal{L}\rangle &= -i\sum_n n \phi^*_n(t) \dot{\phi}_n(t) = -\frac{i}{2}\partial_tC(t) \\
		\langle \mathcal{K} \rangle & = \sum_n n |\phi_n|^2 = C(t) \\
		\langle \mathcal{L} \rangle & = - i\sum_n \phi_n^*(t) \dot{\phi}_n(t) = -\frac{i}{2}\partial_t P(t)
	\end{split}
\end{equation}
In term of renormalized operators, this reads 
\begin{equation}
	\begin{split}
		\langle  \Tilde{\mathcal{K}}  \Tilde{\mathcal{L}}\rangle & = -\frac{i}{2(P(t))^2}\partial_tC(t) \\
		\langle \Tilde{\mathcal{K}} \rangle & = \frac{C(t)}{P(t)} \equiv \Tilde{C}(t) \\
		\langle \Tilde{\mathcal{L}} \rangle & = -\frac{i}{2P(t)}\partial_t P(t)
	\end{split}
\end{equation}
where $\Tilde{C}(t)$ is referred to as renormalized K-complexity. By incorporating all the previously obtained results, the uncertainty equation in Eq.\,\eqref{eqsup:uncertainityrelation} can be reformulated as the dispersion bound on the growth of Krylov complexity in the presence of interactions with the environment, 
\begin{equation}
	\left|\left(\partial_tP(t) \cdot C(t) - \partial_t C(t) \right)\right|^2 \leq 4|b_1|^2 \cdot (P(t))^2\langle (\Delta \Tilde{\mathcal{K}}^\dagger)^2 \rangle\, .
\end{equation}
In term of renormalized K-complexity, we can recast the bound as
\begin{equation}
	\left|(1-P(t))\cdot \partial_t P(t)\cdot  \Tilde{C}(t) + P(t) \cdot \partial_t \Tilde{C}(t)\right|^2 \leq 4|b_1|^2 \cdot (P(t))^2\langle (\Delta \Tilde{\mathcal{K}}^\dagger)^2 \rangle \,.
\end{equation}

\vspace{5mm}
\noindent \textit{On $\langle (\Delta \mathcal{L})^2\rangle $} --- Here in the right hand side of the inequality, we make use of the fact that 
\begin{equation}
	\langle (\Delta \mathcal{L} \left(t=0\right))^2\rangle \geq \langle (\Delta \mathcal{L} \left(t\right))^2\rangle,
\end{equation}
as each of such un-normalized expectation value for a dissipative open quantum system would go through a decay in time due to decoherence. Hence, in the following, we consider the the $\langle (\Delta \mathcal{L})^2\rangle$ expectation value at $t = 0$, 
\begin{align*}
	\mathcal{L} |O(0)\rangle &= \mathcal{L} |p_0\rangle = a_0 |p_0\rangle + c_1 |p_1\rangle \\
	\langle O(0)| \mathcal{L}^\dagger &= \langle q_0| \mathcal{L}^\dagger = a_0 \langle q_0| + b_{1}\langle q_1| \\
	\langle \mathcal{L}^\dagger \mathcal{L} \rangle &= a_0^2 + b_1c_1 \\
	\langle \mathcal{L}^\dagger \rangle &= a_0 \\
	\langle \mathcal{L}\rangle &= a_0 
\end{align*}
so that 
\begin{equation}
	\langle (\Delta \mathcal{L} )^2\rangle = a_0^2 + b_1c_1 - a_0^2  = b_1c_1
\end{equation}
Assuming $b_1 = c_1 = |b_1|$, we have $\langle (\Delta \mathcal{L})^2 \rangle = |b_1|^2$.

\end{document}